\documentclass[prl,twocolumn,showpacs,superscriptaddress]{revtex4-1}

\usepackage{amsmath,amssymb}
\usepackage{xcolor}
\usepackage{graphicx}
\usepackage{braket}
\usepackage[utf8]{inputenc}
\usepackage{bm}

\begin{document}
\date{\today}
\title{Supersymmetric polarization anomaly in photonic discrete-time quantum walks}
\author{Sonja Barkhofen}
\affiliation{Applied Physics, University of Paderborn, Warburger Strasse 100, 33098 Paderborn, Germany}
\author{Lennart Lorz}
\affiliation{Applied Physics, University of Paderborn, Warburger Strasse 100, 33098 Paderborn, Germany}
\author{Thomas Nitsche}
\affiliation{Applied Physics, University of Paderborn, Warburger Strasse 100, 33098 Paderborn, Germany}
\author{Christine Silberhorn}
\affiliation{Applied Physics, University of Paderborn, Warburger Strasse 100, 33098 Paderborn, Germany}
\author{Henning Schomerus}
\affiliation{Department of Physics, Lancaster University, Lancaster, LA1 4YB, United Kingdom}

\begin{abstract}
Quantum anomalies lead to finite expectation values that defy the apparent symmetries of a system.
 These anomalies are at the heart of topological effects in electronic, photonic and atomic systems, where they result in a unique response to external fields but generally escape a more direct observation.
 Here, we implement an optical-network realization of a discrete-time quantum walk, where such an anomaly can be observed directly in the unique circular polarization of a topological midgap state.
 We base the system on a single-step protocol overcoming the experimental infeasibility of earlier multi-step protocols.
 The evolution combines a chiral symmetry with a previously unexplored unitary version of supersymmetry.
 Having experimental access to the position and the coin state of the walker, we perform a full polarization tomography and provide evidence for the predicted anomaly of the midgap states.
 This approach opens the prospect to dynamically distil topological states for quantum information applications.
\end{abstract}

\pacs{
03.67.Ac,
42.50.-p,
03.65.Vf}

\maketitle

\emph{Introduction.---}
Quantum anomalies take a privileged position amongst fundamental physics as they equip quantum systems with robust topological effects. The historic backdrop for quantum anomalies is provided by the Atiyah-Singer index theorem for the Dirac operator \cite{atiyah_index_1963}, which states that the difference of zero modes with positive and negative chirality is a topological invariant. These zero modes are of fundamental significance not only because of their robustness against smooth deformations, but also since their definite chirality defies an apparent symmetry of the system, which results in an anomalous response to symmetry-breaking external fields. An early practical realization is the Su-Schrieffer-Heeger model for polyacetylene \cite{su_solitons_1979}, where the anomalous properties of a midgap state result in charge fractionalization and spin-charge separation \cite{jackiw_solitons_1976}. Interest in this phenomenon therefore quickly transcended the original setting of continuum and lattice field theories \cite{callan_anomalies_1985}, and presently provides a major motivation for research particularly in electronic \cite{nielsen_adler-bell-jackiw_1983,haldane_model_1988, hasan_colloquium:_2010, qi_topological_2011}, superconducting \cite{qi_topological_2011, beenakker_search_2013, teo_topological_2010, beenakker_random-matrix_2015}, photonic \cite{haldane_possible_2008, wang_observation_2009, hafezi_robust_2011, regensburger_paritytime_2012,khanikaev_photonic_2013, regensburger_observation_2013, hafezi_imaging_2013, schomerus_topologically_2013, rechtsman_photonic_2013, lu_topological_2014,poli_selective_2015, zeuner_observation_2015,  mittal_measurement_2016, cardano_detection_2017}
 and ultracold atomic \cite{ jotzu_experimental_2014,mancini_observation_2015,stuhl_visualizing_2015,flaschner_experimental_2016,goldman_topological_2016} systems. In all these settings, zero-modes represent symmetry-protected midgap states with unique finite expectation values of a relevant symmetry operator, resulting in a distinct response when probed by suitable external fields. This includes the formation of anomalous currents, as recently observed in Dirac and Weyl semimetals \cite{xiong_evidence_2015, gooth_experimental_2017}.
An equally early development was the relation of such anomalous behaviour to supersymmetry. In this case systems appear with partners that differ in the number of zero modes, with the prime example being a Dirac particle exposed to a magnetic field \cite{jackiw_fractional_1984,thaller_dirac_2013}. This feature is central to field-theoretic descriptions, but has been much less inquired in practical systems.

In this work we exploit this link via a previously unexplored variant of supersymmetry for the time-evolution operator, and achieve the direct observation of the anomalous expectation value of a zero mode, without the need of an external probe,  in a topological discrete-time quantum walk (QW) \cite{kitagawa_exploring_2010,kitagawa_observation_2012, cedzich_topological_2018, asboth_symmetries_2012, asboth_bulk-boundary_2013, tarasinski_scattering_2014, cardano_statistical_2016, barkhofen_measuring_2017, xiao2017observation, flurin_observing_2017, ramasesh_direct_2017, wimmer_experimental_2017} implemented by a weak coherent laser pulse propagating in a time-multiplexing optical fibre network \cite{schreiber_photons_2010, nitsche_quantum_2016}.
In contrast to proposed and experimentally realised split-step and multi-step protocols in coined QWs \cite{kitagawa_exploring_2010,kitagawa_observation_2012, cedzich_topological_2018, asboth_symmetries_2012, asboth_bulk-boundary_2013, tarasinski_scattering_2014, cardano_statistical_2016, barkhofen_measuring_2017, xiao2017observation, flurin_observing_2017,ramasesh_direct_2017} involving two or more experimental step operations to implement one application of the quantum walk unitary, our protocol exhibits a single step dynamic in which each experimental step directly corresponds to one step of the protocol, which is favourable in terms of losses, resource management and scalability.
The combination of chiral symmetry with supersymmetry results in a topologically non-trivial gapped bandstructure exhibiting four symmetric bands along the quasienergy circle, revealing a topological structure on a three-dimensional torus. These topological features directly relate to an internal degree of freedom, the coin-state of the random walker, which is embodied in the polarization of the laser pulses. While in a suitable basis states originating from the bands exhibit linear polarization, a system with an interface of two topologically distinct systems also contains midgap states whose polarization turns out to be circular. This is the direct manifestation of the anomaly in question. We observe this effect experimentally by performing polarization tomography of the localised output state, as well as by altering the overlap of the input and the midgap state via polarization control.

\begin{figure*}[t]
	\includegraphics[width=0.85\textwidth]{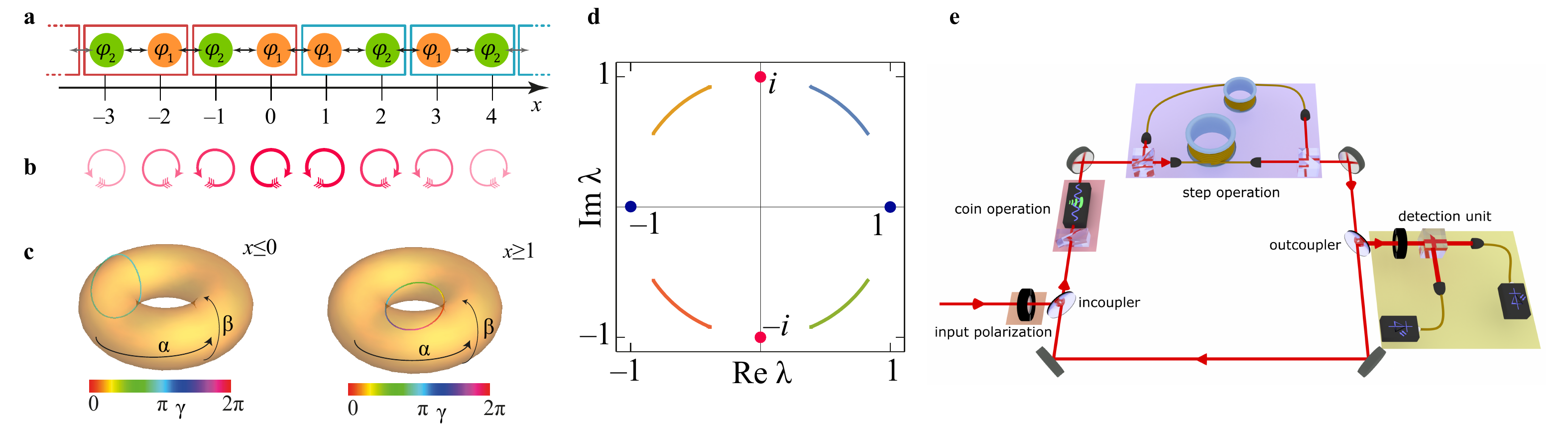}
	\caption{\textbf{Supersymmetric single-step quantum walk} realizing an interface between two topologically distinct phases. (a) Coin structure in the interface configuration, where each disk represents the action of a coin that rotates the polarization by the denoted angles $\varphi_1$ or $\varphi_2$. Across the interface the positions of these coins in the unit cells  (red and blue boxes) are interchanged. (b) Alternating circular polarization of the spatially localised midgap states trapped by the interface. The fading of the color strength away from the interface indicates the intensity decay of the localized midgap state. All extended states display a linear polarization (not shown). (c) Winding of states from the bands around the three-dimensional torus $(\alpha,\beta,\gamma)$, revealing the topological structure of the supersymmetric quantum walk on both sides of the interface. (d) Quasi-energy band structure $\lambda(k) = \exp(-i\epsilon(k))$ comprising four symmetric bands (colored arcs, here shown for $\varphi_1=1$, $\varphi_2=0.2$).  We realize the midgap states pinned to $\lambda = \pm i$ (red dots). (e) Experimental setup using a time-multiplexing optical fibre-loop; see text for details.}
	\label{fig:fig1}
\end{figure*}

\emph{Single step quantum walk protocol.---}
The quantum walk protocol and its experimental realization are illustrated in Fig.~\ref{fig:fig1}. The state
\begin{equation}
|\Psi_t\rangle = \sum_{x\in \mathbb{Z}, c \in\{H,V\}} \Psi_{x,c}(t)|x,c\rangle
\end{equation}
of the quantum walker is defined by the discrete positions $x$ and the coin state $c$, which in our experiments is realized via a train of weak coherent laser pulses and their polarization (H for horizontal, V for vertical). The initial pulse is spatially localised on site $x=1$ with a preset input polarization. This state changes over a time step via the application of position-dependent coin operation
\begin{equation}
\hat{C}(\varphi_x) = \sum_x |x\rangle \langle x| \otimes \begin{pmatrix}
\cos(\varphi_x) &-i\sin(\varphi_x)\\
-i\sin(\varphi_x) & \cos(\varphi_x)
\end{pmatrix}
\end{equation}
rotating the polarization in the H/V basis, followed by a step-operation
\begin{equation}
\begin{aligned}
\hat{S} = &\sum_{x}
\big( \ket{x+1}\!\bra{x}\otimes \ket{H}\!\bra{H}
+ \ket{x-1}\!\bra{x}\otimes \ket{V}\!\bra{V} \big)
\end{aligned}
\label{eq:stepoperator}
\end{equation}
resulting in a unitary evolution governed by $U = \hat{S}\hat{C}(\varphi_x)$.
In the following we consider the bulk configuration, in which the coin angles $\varphi_1$ and  $\varphi_2$ are alternately applied from site to site, and the interface configuration in which a semi-infinite chain with alternating  $\varphi_1$ and  $\varphi_2$ is connected at $x = 1$ to a  chain with alternating  $\varphi_2$ and  $\varphi_1$ (see Fig.~\ref{fig:fig1}a).

\emph{Supersymmetry in quantum walks.---}
We first identify the hidden supersymmetry in the quantum walk, and then use this to predict the anomalous properties of the zero mode in the interface configuration. As typical in the study of topological systems, the key is to connect the features of the zero mode to symmetry constraints of the infinitely periodic bulk system, which we here cast in terms of a unitary variant of supersymmetry that leads to an enlarged set of topological winding numbers.

Previous work considered the bulk system to be periodic after two round trips, so that each wave packet has visited both coins. The hidden symmetry becomes apparent when we consider a single round trip, but follow the amplitudes in a two-site unit cells (blue in Fig.~\ref{fig:fig1}a), where each site carries two polarizations.
Applying Floquet-Bloch theory \cite{kitagawa_exploring_2010, asboth_symmetries_2012,suppmat}, this gives rise to a 4-dimensional unitary evolution
parameterized by a wave number $k$, which is of the explicit form
\begin{eqnarray}
\label{eq:EVequations}
&&u(k) = \begin{pmatrix}
0 & \sigma_xf_{-k}\sigma_x\hat{C}(\varphi_2)\\
f_k \hat{C}(\varphi_1 )& 0
\end{pmatrix} \equiv\begin{pmatrix}
0 & u_{12}(k)\\
u_{21}(k)& 0
\end{pmatrix},
 \nonumber\\
&&f_k = \begin{pmatrix}
1 & 0\\
0& \exp(ik)
\end{pmatrix}.
\end{eqnarray}
Here the blocks (with  Pauli matrix $\sigma_x$) operate on the polarization degree of freedom on a given site.

The bulk bands $\Psi(k)$ are stationary under the application of this evolution,
$u(k)\Psi(k) = \lambda(k)\Psi(k)$, where  $\lambda(k)=\exp(-i\epsilon(k))$ is a propagation factor that can be cast in terms of quasi-energies $\epsilon(k)$. These quasi-energies play the role of the band structure known from autonomous settings, but are to be taken modulo $2\pi$. For the Floquet-Bloch operator \eqref{eq:EVequations} the bands are determined by
the condition $\mathrm{Re} [\lambda^2 (k)]=\cos(\varphi_1 )  \cos(\varphi_2 )  \cos(k)-\sin(\varphi_1 )  \sin(\varphi_2)$.
A sample bandstructure, folded around the unit circle, is shown in Fig.~\ref{fig:fig1}d.
We note that the four bands are related by $\lambda_1 (k)= \lambda_2^* (k) =-\lambda_3 (k)= -\lambda_4^* (k)$, and  separated by gaps at $\lambda=\pm 1$ and $\lambda=\pm i$.

It is clear that these bulk features should arise from general properties of the system. Their topological origin becomes manifest in the symmetric basis
\begin{eqnarray} \label{eq:basistransform}
|H^\prime \rangle &=& \cos(\varphi/2)|H \rangle +i \sin(\varphi/2)|V \rangle \\ \nonumber
|V^\prime \rangle &=& \cos(\varphi/2)|V \rangle +i \sin(\varphi/2)|H \rangle
\end{eqnarray}
in which the Floquet-Bloch operator reads
\begin{equation}
u^\prime(k) = \begin{pmatrix}
0 & \hat{C}(\frac{\varphi_1}{2})\sigma_xf_{-k}\sigma_x\hat{C}(\frac{\varphi_2}{2}) \\
\hat{C}(\frac{\varphi_2}{2})f_{k}\hat{C}(\frac{\varphi_1}{2}) & 0
\end{pmatrix}.
\end{equation}
This displays the two symmetries $u^{\prime \dagger}(k) = \sigma_y u^\prime(k) \sigma_y$, where the Pauli matrix $\sigma_y$ operates on the polarization degrees of freedom, as well as $u^\prime(k) = -\Sigma_z u^\prime(k)\Sigma_z$, where the Pauli matrix $\Sigma_z$ operates on the two positions in the unit cell \cite{suppmat}.
The symmetry induced by $\sigma_y$ constitutes a conventional chiral symmetry for a Floquet operator \cite{kitagawa_exploring_2010, asboth_symmetries_2012} and constraints its eigenvalues to occur in pairs $(\lambda,\lambda^*)$, hence quasienergies $(\epsilon,-\epsilon)$, protecting the gaps at $\lambda=\pm 1$.
The additional symmetry induced by $\Sigma_z$ constraints eigenvalues to occur in pairs $(\lambda,-\lambda)$, hence quasienergies $(\epsilon,\epsilon+\pi)$, and does not have a counterpart in previous investigations.

To identify its origin, we notice that according to
\begin{equation}
u^2(k) = \begin{pmatrix}
u_{12}(k)u_{21}(k) & 0 \\
0 & u_{21}(k)u_{12}(k)
\end{pmatrix}
\end{equation}
upon iteration the Floquet-Bloch evolution (\ref{eq:EVequations}) separates into two partner problems $u_{12}(k)u_{21}(k)$ and $u_{21}(k)u_{12}(k)$, which happen to recover the previously employed split-step protocols \cite{kitagawa_exploring_2010, asboth_symmetries_2012, kitagawa_observation_2012}.
This reduction of a problem with symmetries into two partner problems provides a unitary analogy to the concept of supersymmetry for autonomous Hamiltonians  of the form \cite{jackiw_fractional_1984, thaller_dirac_2013,COOPER1995267,suppmat}
\begin{equation}
H=\begin{pmatrix}
0 & A^\dagger \\
A &0
\end{pmatrix}
~~\mathrm{and~hence}~~~
H^2=\begin{pmatrix}
A^\dagger A & 0 \\
0 &AA^\dagger
\end{pmatrix},
\label{eq:susyh}
\end{equation}
where $A^\dagger A$ and $AA^\dagger$ represent the supersymmetric partners \footnote{For other recent supersymmetric factorizations of discrete autonomous Hamiltonians with relevance to optical systems see Refs.~\cite{PhysRevLett.110.233902,Heinrich2014}.}.
In this light we will call the symmetry induced by $\Sigma_z$ \emph{unitary supersymmetry}.

\emph{Ramifications.---}
While for Hamiltonians of the  form \eqref{eq:susyh} the constraint $\Sigma_zH\Sigma_z=-H$ coincides with a chiral symmetry, in the
Floquet setting the constraints induced by chiral symmetry and unitary supersymmetry are independent and inequivalent, and in combination protect the gaps at $\lambda=\pm i$.
In consequence, the two partner problems exhibit the same spectrum; however, they constitute topologically distinct phases as they are separated by transitions where the gaps at $\lambda=\pm i$ close.

The topological distinction can be asserted by translating these spectral constraints to constraints on the bulk wavefunctions. For our study of particular relevance is the condition
$\langle\Sigma_z \sigma_y \rangle=0$ unless $\lambda=\pm i$, which follows from
\begin{equation}
0= \psi^\dagger(\Sigma_z\sigma_yu^\prime+u^{\prime \dagger}\Sigma_z\sigma_y)\psi = (\lambda+\lambda^{-1})\psi^\dagger\Sigma_z\sigma_y\psi.
\end{equation}
By similar arguments we can
derive the conditions $\langle\sigma_y\rangle=\langle \Sigma_z\rangle=0$, which generally apply when $\lambda\neq \pm i, \pm 1$ \cite{suppmat}.
Physically, the symmetry constraints $\langle\sigma_y\rangle=\langle \Sigma_z \sigma_y\rangle=0$  imply a linear polarization of the bulk Bloch states in the H$^\prime/$V$^\prime$  basis. Mathematically, these conditions  confine the states to geometrically wind around a three-dimensional torus defined by three angles ($\alpha,\beta,\gamma$),
\begin{align} \nonumber
(\cos (\alpha), \sin(\alpha)) &= (\langle \sigma_x(1+\Sigma_z)\rangle, \langle \sigma_z(1+\Sigma_z)\rangle), \\
(\cos (\beta), \sin(\beta)) &= (\langle \sigma_x(1-\Sigma_z)\rangle, \langle \sigma_z(1-\Sigma_z)\rangle), \\ \nonumber
(\cos (\gamma), \sin(\gamma)) &= (\langle \Sigma_x(1-\sigma_y)\rangle, \langle \Sigma_y(1-\sigma_y)\rangle),
\end{align}
as shown in Fig.~\ref{fig:fig1}c.

In the interface configuration, two regions with incompatible winding topology are joined together \footnote{Note that as in other settings the winding numbers depend on the choice of the unit cell \cite{asboth2016short}. Importantly,  the winding numbers on both sides of our interface configuration always differ as long as the unit cell is chosen consistently (as indicated in Fig.~\ref{fig:fig1}a), which is required for the application of the bulk-boundary principle.}.
Applying the bulk-boundary principle \cite{asboth_bulk-boundary_2013,asboth2016short,suppmat}, the interface configuration is then guaranteed to supplement the extended bulk states by spatially confined midgap states,
which furthermore are expected to display anomalous finite expectation values of the relevant symmetry operators.
In our setting, this results in a pair of midgap states pinned to $\lambda =\pm i$ with finite $\langle\Sigma_z \sigma_y \rangle=-1$, which thus display
with an anomalous finite circular polarization that alternates from site to site (see Fig.~\ref{fig:fig1}b). This is the polarization anomaly that we now set out to detect experimentally.

\begin{figure}[t]
	\includegraphics[width=\columnwidth]{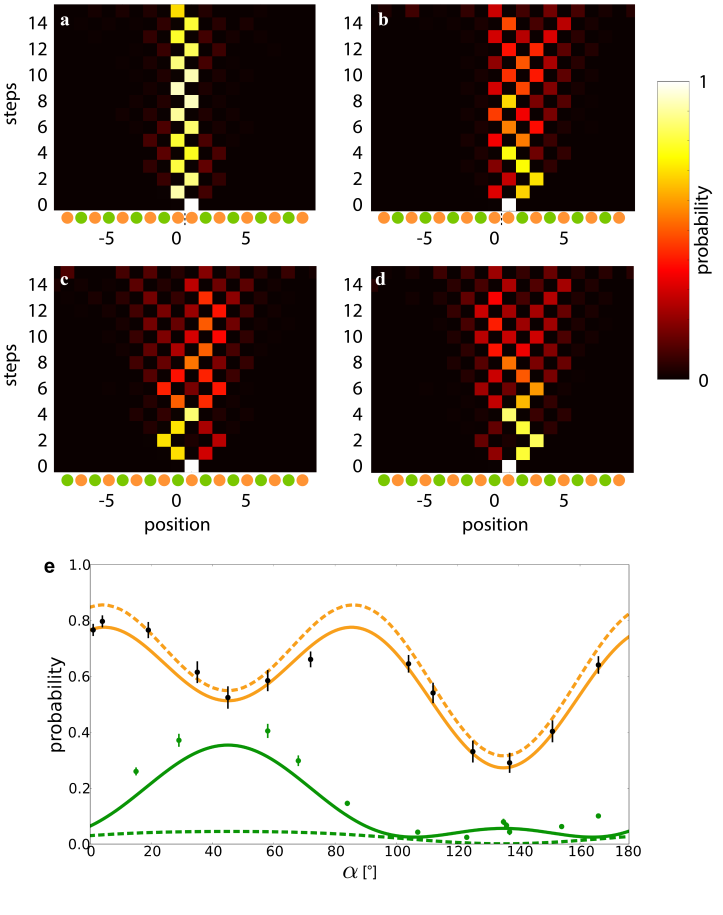}
	\caption{\textbf{Light trapping} for the interface configuration (a,b) compared to the interface-free bulk system (c,d). The exemplary input polarizations are $|H\rangle$ in \textbf{a} and \textbf{c}, $C_\mathrm{QWP}(137^\circ)|H\rangle$ in \textbf{b} and $C_\mathrm{HWP}(50^\circ)|H\rangle$ in \textbf{d} as defined in eqs.~(S17) and (S18). The dependence of the trapped light intensity on the initial polarization is further characterized in \textbf{e} for interface (orange lines, black dots) and bulk (green lines and symbols) configuration. It shows the total intensity after step 13 at position 0 as a function of the initial polarization set by the angle $\alpha$ of the QWP in front of the incoupler (vertical ticks indicating error bars: experimental data; continuous curves: numerical prediction for 13 step; dashed curve: numerical prediction for 100 steps).}
	\label{fig:fig2}
\end{figure}

\emph{Experimental implementation.---}
In the experiments (see Fig.~\ref{fig:fig1}e), the position-dependent coin operations are realized by a Soleil-Babinet compensator (SBC) and a fast switching electro-optic modulator (EOM, red shaded area) \cite{elster_quantum_2015, nitsche_quantum_2016, barkhofen_measuring_2017}.
The shift operation is performed in the well-established time-multiplexing scheme by splitting up the two polarization components at a polarizing beam splitter \mbox{(PBS)} and routing them through fibres of different lengths (blue shaded area) \cite{schreiber_photons_2010, nitsche_quantum_2016}.
The outcoupled pulses are measured with avalanche photodiodes (APDs) in the three bases (H/V, diagonal and circular), giving access to the complete polarization state at each site of the walk.
This detection scheme enables us to observe the polarization-resolved time evolution of the walker and perform a full polarization tomography of the midgap state \cite{suppmat}.

\emph{Results: Light trapping in interface and bulk.---}
We compare a bulk configuration, in which the coin angles alternate between the values $\varphi_1=1.29$, $\varphi_2=0.17$, with an interface configuration, in which the coins are interchanged in half of the system (see Fig.~\ref{fig:fig1}a).
The bulk configuration only supports spatially extended states, which are organised in quasienergy bands $\lambda(k)=\exp(-i\epsilon(k))$ (see Fig.~\ref{fig:fig1}d).
However, in the interface configuration there additionally exist midgap states pinned to  $\lambda=\pm i$, which are spatially localized around the interface.
In the experiments, the difference between the bulk and interface configurations is analysed in detail in Fig.~\ref{fig:fig2}.
Here, we compare the two configurations for different input polarizations of the initial excitation at $x=1$, and study how it spreads over the system.
The difference between both systems is immediately visible.
The midgap state, which we expect to be centred at the interface between sites $x=0$ and $1$,  can trap the initial wave packet (see panels a,b).
This effect displays a strong polarization dependence, and is particularly pronounced for H input polarization.
In contrast, the bulk configuration (c,d) traps a much smaller amount of light, which displays a much weaker polarization dependence.
The polarization dependence is further quantified in panel (e).
Here, we record the detection probability of the quantum walker after 13 steps at the $x=0$ position while varying the angle of a quarter waveplate (QWP) in front of the incoupler.
For the interface system large variations of the trapped light component can be observed, ranging from below 0.3 up to 0.82 (black symbols).
The experimentally observed polarization dependence agrees well with the results of numerical simulations (solid orange curve), which model the quantum walk in detail \cite{suppmat}.
In the bulk system (green symbols and curves) the range of the polarization-dependent variations is much less pronounced.
We extrapolate these results to large step numbers numerically (dashed curves), where a pronounced polarization dependence only remains for the interface configuration.
We also analysed the position dependence of the trapping when exciting the walk not directly at the interface, but scan different input positions (see Fig.~S2 in \cite{suppmat}).
For the polarisation resolved probability histograms demonstrating the spatial localisation of the midgap state see Fig.~S1 in \cite{suppmat}.
These observations uncover a strong and characteristic polarization dependence of the excitability of the midgap state.

\begin{figure}[t]
	\includegraphics[width=\columnwidth]{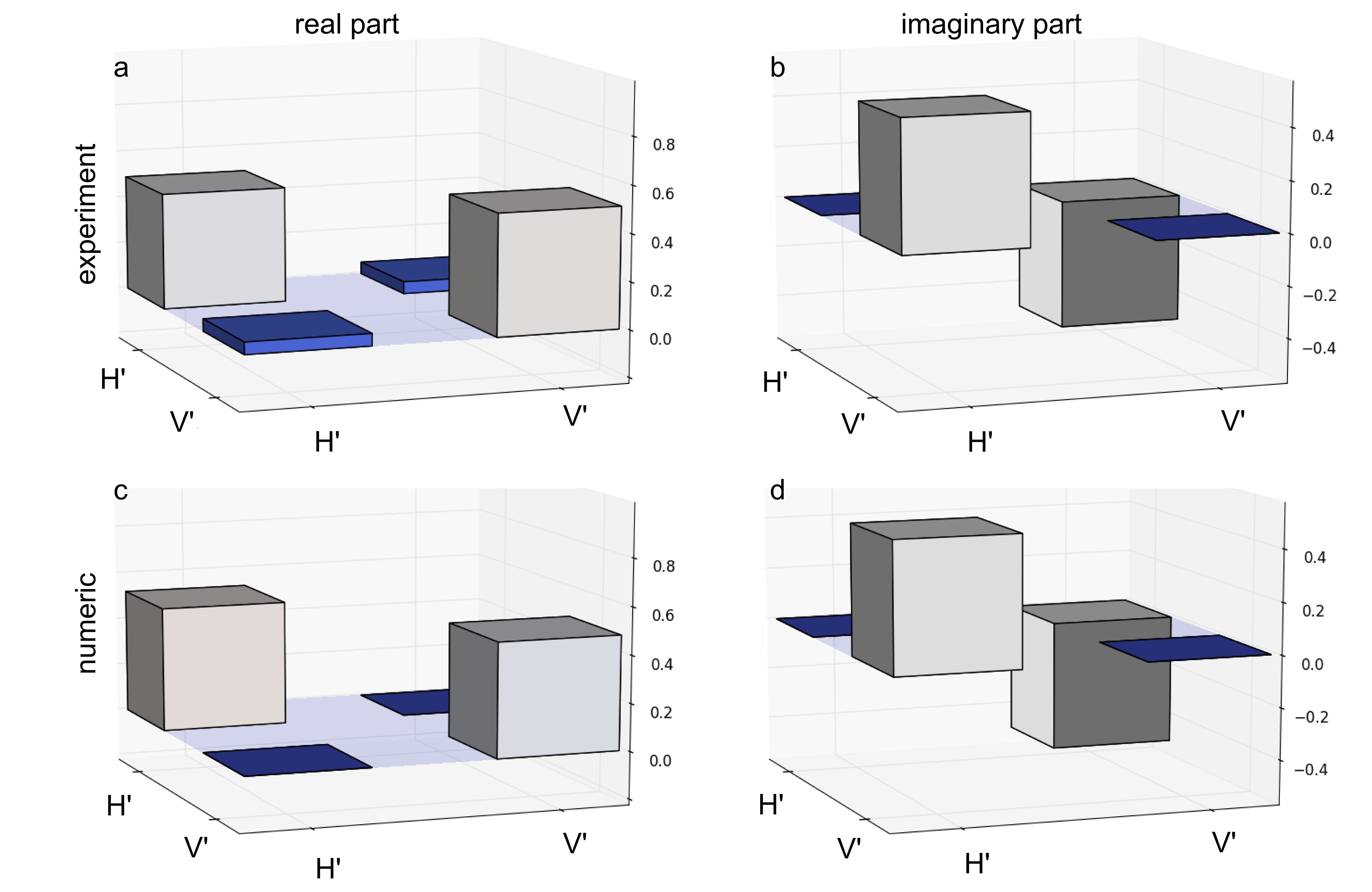}
	\caption{\textbf{Anomalous polarization of the trapped midgap state} from tomography of the polarization state in the interface configuration after step 17 at $x=0$. Note that due to the strong spatial localisation the other positions are hardly occupied.
		The reconstructed complex density matrix from the experiment (a,b) is compared with the numerical prediction in the H$^\prime / $ V$^\prime$ timeframe (c,d).
		The input polarization is $|H\rangle$.
		We observe an almost equal amplitude for the H$^\prime$ and the V$^\prime$ component on the diagonal elements of the real part, while the off-diagonal elements of the imaginary part clearly display a $\pi/2$ phase shift, corresponding to right-handed circular polarization.
		From the experimental data we find the polarization state $(0.70\pm 0.03)| H^\prime \rangle +(0.71\pm 0.02) \exp((0.47\pm 0.02) i \pi) | V^\prime \rangle$, while numerically $0.72| H^\prime \rangle+0.69 \exp(0.50 i \pi )| V^\prime \rangle$. }
	\label{fig:fig3}
\end{figure}

\emph{Results: Detection of the quantum anomaly.---}
In order to demonstrate the anomalous polarization of the midgap state precisely, we measure the full polarization state of the walker after 17 steps on site $x=0$ by performing a tomographic measurement \cite{suppmat}.
 The experimental data presented in Fig.~\ref{fig:fig3} provides the density matrix of the state $(0.70\pm 0.03)| H^\prime \rangle +(0.71\pm 0.02)\cdot \exp((0.47\pm 0.02) i \pi) | V^\prime \rangle$ at $x=0$, which is in excellent agreement with the expected right-handed circular polarization $\sqrt{1/2} (| H^\prime \rangle+i| V^\prime \rangle)$ on the even sites. Analogously, we find left-handed circular polarization $\sqrt{1/2} (| H^\prime \rangle-i| V^\prime \rangle)$ on the odd sites (see  Fig.~S3 in \cite{suppmat}).
 These results verify the anomalous expectation values directly, without relying on currents induced by symmetry-breaking external fields.

\emph{Discussion.---}
In conclusion, we designed a quantum walk that displays a distinctly polarised midgap state.
This allowed us to directly observe an anomalous feature  of a topological zero mode, a fundamental feature that underpins  topological physics in a wide range of settings. 
In our realization
the midgap state is spatially localized at the interface of two topologically distinct systems and situated in a  quasi-energy band gap that arises from the combination of chiral symmetry and previously unexplored unitary supersymmetry. 
In a suitable basis, this gives rise to a circular polarization of the localized midgap state. 
In contrast the bulk states are linearly polarized and spatially extended. 
We demonstrated how to directly address this midgap state via variation of the input polarization, and characterized it via a full polarization state tomography.
The characteristic polarization serves as an avenue to selectively excite the midgap state, as well as to separate it from other eigenmodes by polarization- controlling elements, which both are useful features for possible classical and quantum information and communication applications.

\begin{acknowledgments}
The Group of Paderborn acknowledges financial support from European Commission with the ERC project QuPoPCoRN (no. 725366) and from the Gottfried Wilhelm Leibniz-Preis (grant number SI1115/3-1).
HS acknowledges support by EPSRC via Programme Grant 	EP/N031776/1.
\end{acknowledgments}

%

\end{document}